# Integrating Enterprise Software Applications with Web Portal Technology

Sergey V. Zykov
ITERA Oil and Gas Company
Moscow, Russia
szykov@itera.ru

**Abstract**

Web-portal based approach can significantly improve the entire corporate information infrastructure. The approach proposed provides for rapid and accurate front-end integration of heterogeneous corporate applications including enterprise resource planning (ERP) systems. Human resources ERP component and multimedia data warehouse implementations are discussed as essential instances.

## 1. Introduction

Frequent priority changes in enterprise development require fast and flexible adaptability of management to rapidly changing market conditions. Such adaptability should be based on strategic software integration and its connection to Internet, especially for comprehensive ERP and huge enterprise warehouses.

During the two recent decades, the data models (DM) and architectures underlying software development process have been changed significantly to support object methodologies and interoperability. Attempts of enterprise application integration have also been undertaken [2,4-6].

The main objectives of the paper are development of integrated data and metadata model, application of the model for integrating heterogeneous corporate database-centered resources, formal approach to building a Web-interface enterprise-level solution and overview of improved software implementation process. Research methods meeting the problem domain specific features are based on a creative synthesis of fundamental statements of lambda calculus [11], categories [1] and semantic networks [9].

The data model introduced provides an open, integrated, problem-oriented, event-driven data and metadata management of dynamic, heterogeneous, weak-structured problem domains in a more adequate way than previously known ones. The model suggested allows to generate system architecture and interface solutions for open, distributed, interoperable environments supporting front-end, multi-purpose data warehousing and Internet availability on the basis of CORBA, UML, business-process reengineering (BPR) and web-service technologies.

## 2. Architecture and Interface Requirements

Specific features of the problem domain require support of dynamic multi-level and multi-alternative assignment-based comprehensive estimation of enterprise activity. Interface requirements set should allow dynamic variation of mandatory input fields, flexible access rights differentiation and non-interruptible data integrity support. In respect of architecture, the system should provide interoperability, expandability, flexible adjustment to problem domain changes and easy data/metadata correction.

## 3. The Integrated Data and Metadata Model

### 3.1 The Data Object Model

Mathematical formalisms existing for problem domains are not fully adequate to dynamics and statics semantics. Moreover, current methods of CASE-and-RAD enterprise applications development do not result in solutions of a wide application range; commercial ERP implementations do not provide sufficient flexibility of heterogeneous data handling. According to research results of the problem domain, a computational DM based on object calculus has been built. The model is an innovative synthesis of finite sequence, category and semantic network theories.

Data objects (DO) of the DM can be represented as follows: DO = < concept, individual, state >, where a *concept* is understood as a collection of functions with the same definition area and the same value range. An *individual* implies an essence selected by a problem domain expert, who indicates the identifying properties. *State* changes simulate dynamics of problem domain individuals.

Compared to research results known as yet, the DM suggested features more adequate dynamics mapping for heterogeneous problem domains. The DM also benefits better support for problem-oriented integrated data management. In architecture and interface aspects the DM provides straightforward iterative design of open, distributed, interoperable enterprise software based on UML and BPR methodologies. As far as implementation part is concerned, multi-repository information processing of heterogeneous problem domains is supported. Thus, front-end data access is provided which is based on event-driven procedures, dynamic SQL and Web-service technologies.

The computational model suggested is based on the two-level conceptualization scheme [13]. *Conceptualization* implies a process of establishing relationship between problem domain concepts.

Individuals h, according to the assigned types T, are united in assignment-dependent collections, thus forming variables of sort $H_T(I) = \{h \mid h : I \to T\}$. This formalism is used to simulate problem domain dynamics.

When fixing data model individuals, uniqueness of individualization of data object d from problem domain D by means of the formula $\Phi$ is required:

$\| Ix \Phi (x) \| i = d \Leftrightarrow \{d\} = \{d \in D \mid \|\Phi(d)\| i = 1\}$.

## 3.2 The Metadata Object Model

Let us introduce a compression principle for the computational data object model

$C = Iy: [D] x : D(y(x) \leftrightarrow \Phi ) = \{x : D \mid \Phi\}$

that allows to apply the model to concepts, individuals and states separately, as well as to data objects as a whole. The suggested computational metadata model expands traditional ER-model [3] by adding the following compression principle:

$x^{j+1} Iz^{j+1}: […[D]…] \forall x^j: […[D]…] (z^{j+1}(x^j) \Phi^j)$,

where $z^{j+1}$, $x^{j+1}$ – metadata predicate characters in relation to level j, $x^j$ – individual and $\Phi^j$ – DO definition language construction of level j.

The integrated model for objects of data, metadata and states is characterized by scalability, expandability, metadata encapsulation and transparent visualization. Expandability, adequacy, neutrality and semantic correctness of the formalism introduced provide problem-oriented software design with adequacy maintenance at every stage of the implementation process.

Semantics of computational model of objects of data, metadata and states can be adequately and uniformly formalized by means of typed λ-calculus, combinatory logic, and semantic network-based scenario description.

## 3.3 Model Application for Enterprise Software Integration

### Enhancing Integrated ERP and Multimedia Warehouse with Uniform Portal Representation

Let us consider a Web-browser as a universal client-side software. Most of users even alien to ERP systems are familiar with it. Web-browser comes ready with most of operation systems, and it is easily installable and customizable.

In case of heterogeneous software integration, web browser seems to be the lowest common denominator and a user-friendly solution. Moreover, for a huge and geographically spread enterprise, it makes a perfect and uniform interface to reach the personalized data from virtually any location.

This is one more solid reason to integrate the enterprise software on the portal basis with this universal interface for inter- and intra-office communication, i.e., for Internet, Intranet and Extranet.

Let us consider the following parameters of client appearance and behavior: data access rights, personal preferences (fonts, color settings, etc.), Web browser settings (links, cache, history, etc.) and data access device (Web TV, PDA, mobile phone, terminal, etc.) profile.

Let us assume that A and B are sets. Let $B^A$ stand for the mapping from A to B: $B^A = \{f \mid f : A \to B\}$.

Let us match $B^A$ with the $\|\circ\|$ evaluation function:

$\|\circ\| = \{f \mid f \; B^A \times A \to B\}$.

Thus, $\|\circ\| = (<f, x>) = f(x)$, so $\|<f, x>\| = f(x)$.

Now let us build the semantics network language model. Let us consider an ordered pair of DO of the form

$L=\langle R,C\rangle$, where $R=\{R_1, R_2, \ldots\}$ is a dyadic predicate symbols set and $C=\{C_1, C_2, \ldots\}$ is a set of constants. Therewith, the atomic formulae of the model suggested correspond to simple frames, and terms denote problem domain individuals. Let us construct a frame evaluation procedure using the introduced evaluation function $||\circ||$.

Now let us consider an example of user profile evaluation procedure based on the suggested data model. Let the F functional denote the most general class of users. Let the assignment s={high resolution graphics, multimedia} account for user specific settings. Let F(s) stand for the set of users, for whom the specific settings are restricted to high resolution graphics and multimedia.

Let the assignment p={registered, unregistered, corporate} account for user registration status. Let F(s)(p) designate the set of users with high graphics and multimedia preferences for whom a registration status is assigned, i.e., those who have already visited the Web and/or Intranet site. For the sake of simplicity and without loss of generality, let us consider that site visitors set referred above as functional F, is dependent upon browser settings (v), data access device type (e), personal preferences and access rights: $F=F((v), (e), \ldots)$. In this case, the formula $||F=F((v), (e), \ldots)||$ indicates a formal procedure that evaluates parameterized functional, the expression $||F=F((v), (e), \ldots)(s)||$ evaluates users with given specific settings (s), and the formal procedure $||F=F((v), (e), \ldots)(s)(p)||$ evaluates users with given specific settings (s) and registration status (p). The introduced functional F can be considered an illustration of computational formalism for parameterized procedure of comprehensive profile evaluation of certain visitors categories (from user groups to individuals).

Let us demonstrate that two-level conceptualization scheme is sufficient for the model adequacy. Let us introduce the following denotations:

$||r|| = \{r_{c.s.}, r_{r.s.}\}$ – specific costs;

$||z|| = \{z_{c.s.}, z_{r.s.}\}$ – segmentation degree (i.e., possibility of splitting users into stable and independent groups);

$||q_i|| = q_i$ – overheads;

$||l_i|| = l_i$ – duration of the request processing stage (download, dynamic form or report creation, etc.);

$||n_i|| = n_i$ – number of request processing stages.

Evaluated values are generalized, i.e., there is no uniqueness of value choice for specific costs and segmentation degree. Generalization level decrease is achieved by considering an assignment point s:

$$||z||(||s||) = \begin{cases} ||z||(\text{higraph}) = z_{\text{higraph}}, \\ ||z||(\text{mmedia}) = z_{\text{mmedia}}; \end{cases}$$

$$||r||(||s||) = \begin{cases} ||r||(\text{higraph}) = r_{\text{higraph}}, \\ ||r||(\text{mmedia}) = r_{\text{mmedia}}. \end{cases}$$

Moreover, further generalization level decrease by introducing the second assignment p does not succeed:

$||z||(||s||)(||p||) = ||z||(||s||);$

$||r||(||s||)(||p||) = ||r||(||s||).$

The result obtained can be explained by the fact that the evaluation procedure involves visitor position in the data access rights policy.

However, it is obvious that overheads $q_i$ are dependent both on user-specific settings functions and on registration status, i.e. we should let $||q_i|| = \{q_{i\text{ higraph}}, q_{i\text{ mmedia}}\}$. The equality $||q_i|| = q_i$ implies that $q_{i\text{ higraph}} = q_{i\text{ mmedia}} = q_i$.

Similarly, switching to a multimedia data warehouse problem domain suggests various data types and scenarios of handling them. In particular, multimedia data profile is dependent upon the category of data source (audio record, video record, static image) and it can split into sub-categories of photos, logos and catalogues for static images.

The data model proposed is equally applicable for both multimedia storage and ERP system problem domains.

## 4. The Integrated ERP Implementation

### 4.1 Customizing the Implementation Scheme

During design process, ERP specification is transformed from problem domain concepts to data model entities, then, further, to DBMS scheme (with *PL\SQL* as DO manipulation language), and, finally, to target ERP description with required architecture and interface. As a result of problem domain analysis, computational DM and generalized scheme of ERP development [16] have been customized to satisfy the required problem domain conditions.

Web applications implementation is finalizing ERP implementation stage. *Oracle Internet Developer Suite* and *Oracle Portal* serve a gateway between data warehouse and corporate Internet and Intranet sites. With this technological enhancement, heterogeneous data sources become interconnected. When content-critical warehouse updates occur, content is automatically updated accordingly. Periodical and manual data updates and retrievals are also supported options.

## 4.2 Problem-Oriented Interface and Event-Driven Architecture

According to the detailed enterprise software design sequence, a comprehensive heterogeneous repository processing scheme is introduced that allows users to interact with distributed database in a certain state depending on dynamically activated (i.e., assigned) scripts. Thus, scripts (in a form of data access profiles and stored object-oriented program language procedures) are initiated depending on user-triggered events. Scripts provide transparent and intellectual client/server front-end user-to-database connection. Dynamically adjustable database access profiles provide high fault tolerance and data security both for ordinary and privileged system users in heterogeneous environment. The profiles are implemented using CORBA technology as an intellectual media between end-user and heterogeneous data warehouses. Depending on semantics-oriented user profile structure, certain database connection and access level profiles are dynamically assigned. The profiles are valid only until the end of data exchange session. According to the hierarchy, users access data under one of the basic scenario profiles. Access is granted not only to data, but also to metadata (i.e., data object dimensions, integrity constraints, access rights, browser parameters, user preferences, multimedia data types etc.).

Administrative users have extended access to metadata. Thus, under the model introduced, data and metadata objects are manipulated uniformly. This makes system interface a problem-oriented, straightforward and uniform one and significantly increases system performance. Portal-enhanced data warehouse processing scheme under conditions of event-driven architecture is presented in fig.1. Client-side web page object states can change depending on event script execution. Though the warehouse data remains unchanged, user can request data update, or produce a query. Moreover, front-end interface itself is also client profile dependent. Options include personal preferences (multimedia data types, color schemes, screen resolution etc.), data access device and Web browser settings.

The essential benefit of event-orientation for the ERP global network extension is that corporate users get additional access to Intranet resources of corporate Internet site. Registered (and/or extranet) users access some extra data compared to non-privileged ones.

Warehouse data access is also dependent on user profile. At the upper level of data access hierarchy, clients can be divided into administrators, managers and ordinary users. Judging by the profile, data and metadata object states (i.e., system interface) are changed. For instance, a web designer rights assignment provides full access to interface elements database, while portal content manager can get full access to another warehouse instantiation by means of a different interface instance.

When implementing multimedia data warehouse it is essential to handle a large number of data formats (video, audio, graphical images) in a uniform and a transparent way for the warehouse server. The so-called cartridge mechanism is used in Oracle database products for this purpose.

A cartridge can represent a class for a complex business object in an application, for which there is no straightforward way of representing it with generic DBMS server data types. A cartridge can be produced by a third-party vendor according to certain business logic. This logic may be implemented in Java, such programming languages as PL/SQL, C, C++ (and even Perl, which is widely used in Web applications) through external procedures mechanism. Cartridges are using a uniform and powerful IDL (Interface Definition Language) interface, thus making heterogeneous components integration possible under CORBA distributed objects technology.

The pilot implementation is using standard multimedia cartridge that helps to organize a uniform heterogeneous data warehouse and to perform a fast native-format problem-oriented search of video fragments, audio records, photos and static images.

Scenario-based end-user interface results in higher degree of interactivity, user-friendliness and security. User profiles (i.e. assignments) can be stored in metadata base of visitors, and, depending on their properties, data access and representation level could be customized. Sets of Web pages accessible and their content are dependent on user profile. Client-side profile also accounts for preferable information layout depending both on personal preferences and on the device type for which the data is customized. User profile and preferences are also vital for client analysis, and resulting statistic reports serve a foundation for performance and interface optimization.

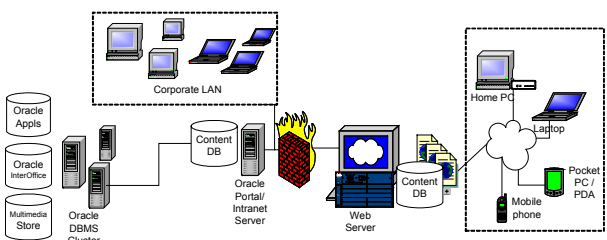

**Figure 1 Enterprise applications integration approach based on portal technology**

Integrating Enterprise Software Applications with Web Portal Technology

## 4.3 Implementation Description

The introduced methodology has been practically approved during ERP system improvement at *ITERA* International Group of Companies. Attempt has been made to cement the environment of *Oracle Applications* family of enterprise-level financial, commodity and document management systems with the uniform integrated portal interface.

From the system architecture viewpoint, the integrated ERP provides certain level of data input, correction, analysis and output depending on front-end position (i.e., assignment) in user hierarchy. Interactive interface is represented by portal-based problem-oriented form designer, report generator, on-line documentation and administration. The enhanced ERP system database supports the integrated storage for data (i.e., information for on-line users) and metadata (i.e., data object dimensions, integrity constraints and business process parameters). During the enhanced ERP design process, problem domain DM specification (in a semantic network representation) is transformed into use-case UML diagrams, then, by means of *Oracle Developer/2000* and integrated CASE-tool, - into ER-scheme and, finally, into the attributes of target databases. *Oracle Internet Developer Suite* and *Oracle Portal* have been used to transform the ERP components into a uniform and consistent Web-wired portal application.

On the basis of the information model developed, an architecture and interface solution for integrated HR management software has been designed. To prove adequacy of the model developed and component integration algorithm suggested, software prototype has been designed. To provide required levels of industrial scalability and fault tolerance, judging by the results of exhaustive CASE-and-RAD software analysis, *Oracle Developer/2000* toolkit has been chosen as a solution supporting UML and BPR methodologies. According to specification requirements developed by the author, implementation had been significantly improved by extracting essential information from integrated ERP and publishing it in the corporate Internet and Intranet portal.

The portal is accumulating information from various ERP problem-oriented modules. For example, the HR subsystem provides a number of significant data items for corporation profile portal pages including total establishment, number of countries and companies that represent the corporation. Since the integrated solution implemented includes vacancy module as a part of its HR subsystem, dynamically updated vacancy data Internet page can be easily produced. Similarly, financial components could provide data for a number of periodical or privileged user-triggered financial reports. Data examples include revenues, profits, production dynamics, stock values, etc. Same as with HR subsystem, dynamics tracking is important. Corporation development plans based on deferred charges could be published. Production manufacturing module can provide productivity and capacity data for executive summaries and company profile Web and Intranet pages. Address book from document control subsystem serves for contact information and provides automatic feedback routing through corporate organizational structure. For further enhancement of corporate Web site performance and interface, data published in the Web page is dynamically updated by an event-driven software agent.

The integrated ERP data warehouse is stored in the data center. ITERA Group warehouse is based on *Oracle Financials, Human Resource*, multimedia DBMS cartridge and *Portal*. The full-scale implementation is based on hardware platform of an IBM RS/6000 two-server high availability cluster running under IBM AIX operating system. The system itself has passed a four-year test in a large corporation, while the portal solution is in progress of enterprise-wide implementation.

## 5. Results and Conclusion

A computational data model has been introduced that provides integrated manipulation of data and metadata objects, especially in rapidly changing heterogeneous problem domains. The model is an alloy of methods of finite sequences, categories and semantic networks.

On the basis of the formal model, an original and comprehensive iterative scheme for Web portal-integrated ERP design and implementation has been proposed. The scheme includes an algorithm for new component integration into existing ERP environment. An innovative integration algorithm suggested provides adequacy, consistency and data integrity; algorithm details are presented in [16].

According to the approach suggested, a comprehensive Web-portal ERP interface has been designed. The interface is based on an open and extendable architecture. As a first step towards implementing the enterprise resource management solution, a fast event-driven software prototype has been developed on the basis of the designed UML-based interface and architecture scheme. Using the prototype testing results, a pilot version of the full-scale object-oriented ERP application portal has been designed. The solution is being customized for corporate resource management and implemented at an enterprise with around 1000 employees.

Web portal-based ERP solution is promising significant decrease in time and costs of implementation. Other major benefits include growth of portability, expandability, scalability and ergonomics levels in comparison with existing commercial software of the kind. Iterative multi-level software design scheme is based on formal model unifying object-oriented methods of data (data objects) and knowledge (metadata objects) management. Industrial implementation of the integrated Internet-embracing ERP components has been carried out using integrated CASE, RAD and portal-building tools. Testing experience has proved importance, originality and efficiency of the approach suggested. Theoretical and practical statements outlined in the paper have been approved by successful implementation of the pilot version of full-scale ERP portal solution at *ITERA* International Group of Companies. The author is going to continue research in order to turn the enhanced ERP into an integrated corporate Intranet and e-commerce solution.

## References

1. K. Baclawski, D. Simovici, W. White. "A categorical approach to database semantics". *Mathematical Structures in Computer Sci.*, vol.4, p.p.147-183, 1994
2. D. Calvanese, G. Giacomo, M. Lenzerini, D. Nardi, R. Rosati. "Source Integration in Data Warehousing". *DEXA Workshop* 1998, p.p.192-197
3. E. F. Codd. "Relational Completeness of Data Base Sublanguages Data Base Systems". In: Rustin R. Eds.,.- New York; Prentice Hall, 1972 (*Courant Computer Science Symposia* Series No.6)
4. D. Florescu, A. Y. Levy. "Recent Progress in Data Integration - A Tutorial". *ADBIS* 1998, p.p.1-2
5. Y. Kambayashi. "Research and Development of Advanced Database Systems for Integration of Media and User Environments". *DASFAA* 1999, p.p.3-5
6. D. S. Linthicum. "Enterprise Application Integration". Addison Wesley Longman. ISBN 0-201-61583-5. Nov.1999. 377 p.
7. Information about Oracle software mentioned in this paper available at WWW: http: // www.oracle.com
8. R. Orfali, D. Harkey, J. Edwards. "The Essential Client/Server Survival Guide", 2nd Edition. Wiley Computer Publishing, 1996, 678p.
9. N. D. Roussopulos. "A semantic network model of data bases", Toronto Univ., 1976
10. D. S. Scott. The lattice of flow diagrams, In: *Symposium on Semantics of Algorithmic Languages*.- Berlin, Heidelberg, New York: Springer-Verlag, 1971, p.p.311-372. (*Lecture Notes in Mathematics* No. 188)
11. D. S. Scott. "Relating theories of the λ-calculus". In J.Hinhley and J.Seldin, eds., *To H.B.Curry: Essays on combinatory logic, lambda calculus and formalism*, p.p.403-450. Academic Press, 1980
12. D.S. Scott. "Domains for denotational semantics". In M.Nielsen and E.M.Schmidt, eds., *Lecture Notes in Computer Science*, vol.140, p.p.577-613, Aarhus, Denmark, 12-16 July 1982. Springer
13. V. E. Wolfengagen. "Object-oriented solutions. *ADBIS* 1996, p.p.407-431. Springer
14. V. E. Wolfengagen. "Event Driven Objects". *CSIT'99*. Moscow, Russia, 1999 p.p.88-96
15. S. V. Zykov. "Integrated Human Resources Information Systems: Involving Extra Data Sources Centered around Groupware". *CSIT'99*. Moscow: MEPhI, 1999, p.p. 209-219
16. S. V. Zykov. "Enterprise Resource Planning Systems: the Integrated Approach". *CSIT'2001*, Ufa, Russia. Sept.21-26, 2001.
17. S. V. Zykov. " The Integrated Approach to ERP: Embracing the Web". *CSIT'2002*, Patras, Greece. Sept.18-20, 2002.